\documentstyle[prd,aps,floats]{revtex}
\input{psfig.tex}
\tighten
\begin{document} 
\draft

\preprint{DAMTP/R-97/??}
\title{Second order Lagrangian and symplectic current for gravitationally perturbed Dirac-Goto-Nambu strings and branes} 
\author{Richard A. Battye{$^{1}$} and Brandon Carter{$^{2}$}}
\address{${}^1$ Department of Applied Mathematics and Theoretical
Physics, University of Cambridge, \\ Silver Street, Cambridge CB3 9EW,
U.K. \\ ${}^2$  Department d'Astrophysique Relativiste et de
Cosmologie, Centre National de la Recherche Scientifique,
\\ Observatoire de Paris, 92195 Meudon Cedex, France.  } 
\maketitle
\begin{abstract} We extend a recent analysis of gravitational
perturbations on Dirac-Goto-Nambu type strings, membranes and higher
dimensional branes. In an arbitrary gauge, it is shown that the
relevant first order  equations governing the displacement vector of
the worldsheet  and metric perturbation are obtainable from a variational
principle whose Lagrangian is constructed as a second order
perturbation of the standard Dirac-Goto-Nambu action density. A
symplectic current functional is  obtained as a by-product that is
potentially useful for the derivation of conservation laws in
particular circumstances.  \end{abstract}

\newcommand{\beq}{\begin{equation}} \newcommand{\feq}{\end{equation}}
\newcommand{\bel}{\begin{equation}\label}
\newcommand{\bea}{\begin{eqnarray}} \newcommand{\fea}{\end{eqnarray}}

\def\ma{m_{\rm a}}
\def\fa{f_{\rm a}}
\def\na{n_{\rm a}}
\def\dalemb#1#2{{\vbox{\hrule height .#2pt
\hbox{\vrule width.#2pt height#1pt \kern#1pt\vrule width.#2pt}
\hrule height.#2pt}}}
\def\square{\mathord{\dalemb{5.9}{6}\hbox{\hskip1pt}}}
\def\tdot{\kern -8.5pt {}^{{}^{\hbox{...}}}}
\def\dotprime{\kern -8.0pt{}^{{}^{\hbox{.}~\prime}}}
\def\ov{\overline}
\def\og{\eta} 
\def\ag{\perp}
\def\nabl{\nabla} \def\onab{\ov\nabl}
\def\R{{\cal R}} 
\def\I{{\cal I}} 
\def\L{{\cal L}} \def\Lg{{\cal L}_{_{\rm\bf G}}}
\def\Tg{T_{\!\!_{\rm\bf G}}\!} 
\def\sp{\ \,}

\font\fiverm=cmr5

\def\spose#1{\hbox to 0pt{#1\hss}} 
\def\Libra{\spose {--} {\cal L}}
\def\La{{\cal L}}
\def\Diam{\spose {\raise 0.3pt\hbox{+}} {\diamondsuit}  }

\def\diamL{\spose {\lower 5.0pt\hbox{\ \fiverm L}} {\raise 0.3pt\hbox 
{$\,\diamondsuit$}}  }

\def\gL{\spose {\lower 7.0pt\hbox{\fiverm L} } {g}}
\def\gE{\spose {\lower 7.0pt\hbox{\fiverm E} } {g}}
\def\dL{\spose {\lower 5.0pt\hbox{\fiverm L} } {\delta}}
\def\dE{\spose {\lower 5.0pt\hbox{\fiverm E} } {\delta}}
\def\GL{\spose {\lower 5.0pt\hbox{\fiverm L} } {\Gamma}}
\def\GE{\spose {\lower 5.0pt\hbox{\fiverm E} } {\Gamma}}
\def\RL{\spose {\lower 5.0pt\hbox{\ \fiverm L} } {\cal R}}
\def\RE{\spose {\lower 5.0pt\hbox{\ \fiverm E} } {\cal R}}
\def\nabL{\spose {\lower 6.0pt\hbox{ \fiverm L} } {\nabla}}
\def\nabE{\spose {\lower 6.0pt\hbox{ \fiverm E} } {\nabla}}

\def\A{{\cal I}}
\def\AI{\spose {\raise 3.0pt\hbox{$\ \acute{\ }$} } {\A}}
\def\AII{\spose {\raise 3.0pt\hbox{$\, \acute{\ }\!\acute{\ }$} } {\A}}
\def\AJ{\spose {\raise 3.0pt\hbox{$\ \grave{\ }$} } {\A}}
\def\AIJ{\spose {\raise 3.0pt\hbox{$\acute{\ }\grave{\ }$} } {\A}}
\def\AJJ{\spose {\raise 3.0pt\hbox{$\,\grave{\ }\!\grave{\ }$} } {\A}}
\def\AJI{\spose {\raise 3.0pt\hbox{$\,\grave{\ }\!\acute{\ }$} } {{\cal I}}}

\def\hI{\spose {\raise 3.0pt\hbox{$\ \acute{\ }$} } {h}}
\def\hII{\spose {\raise 3.0pt\hbox{$\, \acute{\ }\!\acute{\ }$} } {h}}

\def\hJ{\spose {\raise 3.0pt\hbox{$\ \grave{\ }$} } {h}}
\def\hIJ{\spose {\raise 3.0pt\hbox{$\acute{\ }\grave{\ }$} } {h}}
\def\hJJ{\spose {\raise 3.0pt\hbox{$\,\grave{\ }\!\grave{\ }$} } {h}}
\def\hJI{\spose {\raise 3.0pt\hbox{$\,\grave{\ }\!\acute{\ }$} } {h}}

\def\xiI{\spose {\raise 3.0pt\hbox{$\ \acute{\ }$}} {\xi}}
\def\xiII{\spose {\raise 3.0pt\hbox{$\, \acute{\ }\!\acute{\ }$}} {\xi}}
\def\xiJ{\spose {\raise 3.0pt\hbox{$\ \grave{\ }$}} {\xi}}
\def\xiJI{\spose {\raise 3.0pt\hbox{$\,\grave{\ }\!\acute{\ }$}} {\xi}}
\def\xiJJ{\spose {\raise 3.0pt\hbox{$\, \grave{\ }\!\grave{\ }$}} {\xi}}
\def\xiIJ{\spose {\raise 3.0pt\hbox{$\acute{\ }\grave{\ }$} } {\xi}}

\def\F{F}
\def\FI{\spose {\raise 3.0pt\hbox{$\ \acute{\ }$} } {\F}}
\def\FJ{\spose {\raise 3.0pt\hbox{$\, \grave{\ }$} } {\F}}

\def\Q{{Q}}
\def\QI{\spose {\raise 3.0pt\hbox{$\ \acute{\ }$} } {\Q}}
\def\QJ{\spose {\raise 3.0pt\hbox{$\, \grave{\ }$} } {\Q}}

\def\U{J}
\def\UJI{\spose {\raise 3.0pt\hbox{$\,\grave{\ }\!\acute{\ }$} } {\U}}
\def\UIJ{\spose {\raise 3.0pt\hbox{$\acute{\ }\grave{\ }$} } {{\U}}}

\def\V{V}
\def\VJI{\spose {\raise 3.0pt\hbox{$\,\grave{\ }\!\acute{\ }$} } {\V}}
\def\VIJ{\spose {\raise 3.0pt\hbox{$\acute{\ }\grave{\ }$} } {\V}}

\def\C{C}
\def\CI{\spose {\raise 3.0pt\hbox{$\ \acute{\ }$} } {\C}}
\def\CJI{\spose {\raise 3.0pt\hbox{$\,\grave{\ }\!\acute{\ }$} } {\C}}
\def\CIJ{\spose {\raise 3.0pt\hbox{$\acute{\ }\grave{\ }$} } {\C}}
\def\Cr{S}
\def\CrI{\spose {\raise 3.0pt\hbox{$\ \acute{\ }$} } {\Cr}}
\def\CrJI{\spose {\raise 3.0pt\hbox{$\,\grave{\ }\!\acute{\ }$} } {\Cr}}
\def\CrIJ{\spose {\raise 3.0pt\hbox{$\acute{\ }\grave{\ }$} } {\Cr}}

\section{Introduction}

The purpose of this work is to extend our recent
analysis~\cite{batcar95} of the dynamics of small perturbations of the
worldsheet for a simple, internally structureless Dirac-Goto-Nambu type
string, membrane, or higher dimensional brane in a pseudo-Riemannian
background of arbitrary dimension. The main motivation for this
investigation is to prepare the way for a study radiation backreaction
that has already been undertaken~\cite{batshell95,batshell96} in the
technically simpler, but physically more speculative context of scalar
axion fields. In view of this ultimate objective, our
analysis~\cite{batcar95} specifically allowed for the effect of
perturbing gravitational waves, unlike earlier work that was restricted
to effectively free perturbations of the worldsheet in a fixed
flat~\cite{garvil93} or curved~\cite{car93,guv93,frola94} space-time
background. The main result of this analysis to date has been the
demonstration~\cite{carbat98} that the divergent part of the
gravitational self interaction exactly cancels for a Nambu-Goto string
in ordinary four-dimensional space-time, contrary to previous claims
in the literature~\cite{CHH90,QS90}, but consistently with concurrent
work~\cite{buodam98}.

Most simple treatments of the dynamics of branes use a formalism
originally developed by Eisenhart~\cite{frola94}, using the explicit
details of a specifically chosen worldsheet reference scheme at each
step. However, particularly when there are other physically
independent fields (gravitational or otherwise) to be taken into
account, it is efficient to use a more economical
treatment~\cite{car93,guv93} in which the amount of auxiliary
mathematical structure involved is reduced. The approach we have
used~\cite{batcar95,car93} reduces the mathematical paraphernalia to
its strict minimum by working exclusively in terms of fields whose
tensorial components are defined directly in terms of the background
space time coordinates, that is,  without any reference to special
frames or internal coordinates. While particular coordinate systems are
genuinely useful for specific applications, the excessive number of
indices can lead to the mistakes which have already been discussed
above.

Although it did not allow for gravitational wave perturbations, the
first of the works discussed above~\cite{car93} went further than our
more recent generalization~\cite{batcar95}. In both cases, the first-order perturbation equations governing the dynamical evolution of the
relevant fields (a vectorial surface field $\xi^{\mu}$ in the case of
ref.~\cite{car93} and the surface field plus the gravitational wave
perturbation $h_{\mu\nu}$ for ref.\cite{batcar95}) were deduced. But
for the simpler case of just the surface field, it was shown that these
dynamical equations could be derived via a variational principle, from
a Lagrangian of quadratic order in $\xi^{\mu}$, which was constructed
as a second order perturbation of the original action --- in this case
the Dirac-Goto-Nambu action. A by-product of this second-order approach
is the construction of a bi-linear symplectic surface current satisfying
a Noether type conservation law.

In the present work it is shown how to construct an equivalent
second-order Lagrangian, and the corresponding bi-linear symplectic
current in the more general case which includes gravitational wave
perturbations. From this Lagrangian, we deduce that the same dynamical
equations of motion already found in ref.~\cite{batcar95} and used in
ref.~\cite{carbat98}. However, it will be seen that the gravitational
waves perturbations act as a source term for the symplectic current, and
hence in this more general case the relevant Noetherian surface current
no longer satisfies a strict conservation law.

We will use a similar notation scheme as before to describe a p-brane
with a $p+1$ dimensional worldsheet, in $n$ space-time dimensions. As
emphasized already, the need to refer any internal worldsheet
coordinates, $\sigma^a$ (a=0, ..., p) say, is avoided by basing the
analysis on the {\it first} fundamental tensor, with components
$\eta^{\mu\nu}$ defined with respect to ordinary background space-time
coordinates $x^\mu=X^\mu(\sigma^a)$ ($\mu$ =1, ..., n). This tensor is
determined by the background space-time metric $g_{\mu\nu}$ as the
projection of the contravariant inverse of the induced metric on the
worldsheet, that is, in terms of the internal coordinates (using
$\partial_a$ for partial differentiation), as
$\eta^{\mu\nu}=\gamma^{ab}\partial_a X^\mu \partial_b X^\nu$, where the
induced metric is given by $\gamma_{ab}=g_{\mu\nu}\partial_a
X^\mu\partial_b X^\nu$.  Contracting any vector with this fundamental
tensor has the effect of projecting it onto the part tangential to the
worldsheet, while the normal part is obtainable by contraction with the
orthogonal projection tensor
$\perp^{\!\mu\nu}=g^{\mu\nu}-\eta^{\mu\nu}$. We denote the operator of
tangiantially projected covariant differentiation --- the only kind
that is well defined for a tensor field with support  confined to the
worldsheet --- by
\bel{covdiff}
\ov\nabla{^\mu}=\eta^{\mu\nu}\nabla_{\nu}\,,
\feq
where $\nabla_{\nu}$ is the usual operator of covariant differentiation
as defined with respect to the Riemannian background connection,
namely
$\Gamma^{\lambda}_{\sp\sigma\tau}=g^{\lambda\rho}(g_{\rho(\sigma,\tau)}
-{_1\over^2}g_{\sigma\tau,\rho})$ using round brackets for index 
symmetrisation. The information characterizing the various 
(intrinsic and extrinsic) kinds of curvature associated with the worldsheet 
embedding is provided~\cite{car96} by the {\it second} fundamental tensor,
as defined in terms the first one by
\bel{secten}
K_{\mu\nu}^{\sp\sp\rho}=\eta_{\sigma\nu}\ov\nabla_{\mu}\eta^{\rho\sigma}\,.
\feq
Since it is automatically surface tangential, as well as symmetric,
with respect to its first two indices but surface orthogonal with
respect to its last index, contraction with the former provides its
only non-identically vanishing trace, namely the curvature vector
\bel{trace}
K^\rho=\eta^{\mu\nu}K_{\mu\nu}^{\sp\sp\rho}\,.
\feq

The non-trivial dynamical requirement that this vector $K^\rho$ should
actually vanish is the tensorially covariant expression of the
equations of motion for the kind of strings and higher branes that we shall be
considering, namely those characterised by the Dirac-Goto-Nambu action
principle. In this case, the relevant action ${\cal I}$ is just proportional
to the geometrically induced measure of the worldsheet, meaning that it is
obtained by taking a {\it fixed} value for the Lagrangian density $\ov\L$
in the general expression for a worldsheet action, as given by
\bel{action}
\A=\int \ov\L d\ov{\mit\Sigma}\,, \feq
where $d\ov{\mit\Sigma}=\Vert\gamma\Vert^{1/2}d^{p+1}\sigma$, in terms of internal coordinates.

\section{Relation between Lagrangian and Eulerian variations}

In order to proceed with the second-order analysis of the action
principle, we need to define the relevant perturbations and in
particular the difference between Eulerian and Lagrangian variations.
In ref.~\cite{batcar95} we only included the very much simpler
first-order perturbations, while in ref.~\cite{car93} second-order
perturbations were included in a fixed background, that is, no Eulerian
variation. In this section, we generalize these second-order variations
to include Eulerian variations of the metric. The starting point for
this is the finite action variation, which takes the form
\bel{expansion}
\Delta\I=\delta I+{_1\over ^2}\delta^2\I+ {\cal O}\{\delta^3\} \,,\feq
where $\delta$ is an infinitesimal variation operator normalised
with respect to a variation parameter with magnitude proportional to
that of the displacement vector field. 

The Eulerian variation of the metric is given by 
\bel{Euler}
g_{\mu\nu}\quad \Rightarrow \quad \gE_{\mu\nu}=g_{\mu\nu}+h_{\mu\nu} \,,
\feq
defined with respect to some predetermined position identification
scheme, for example, harmonic coordinates. It is possible in the
analysis of a smoothly distributed fluid to work entirely with
Eulerian, that is, `fixed point', variations. However, for the cases
under consideration here (point particles, strings, and higher
dimensional branes) where the relevant fields are confined to a lower
dimensional support manifold that may be  displaced by the
perturbation, it is evident that the concept of an Eulerian variation
will generically fail to be well defined, so that it becomes necessary
to use an approach based on Lagrangian, that is, `comoving',
variations. A similar approach is often also used in dealing with
smooth fluid media.

As a prerequisite for this, the relevant Eulerian background
variation (\ref{Euler}) must first be translated into Lagrangian form. It is
well known that the Lagrangian variation
\bel{Delta}
\Delta_{\mu\nu}=\gL_{\mu\nu}-g_{\mu\nu}\,,\feq
representing the difference between the comoving perturbed metric
$\gL_{\mu\nu}$ and the unperturbed value $g_{\mu\nu}$ will deviate from
its Eulerian analogue by a term that to first order in the displacement
$\xi^\mu$ will be given just by the corresponding Lie derivative of the
unperturbed field. In order to go to second (and higher) order it is
necessary to take care in specifying just how the displacement
corresponding to $\xi^\mu$ is defined. The natural way to do
this~\cite{car96} (as suggested, for example, by the Bunting identity
~\cite{B86}, and the analysis of Boisseau and Letelier~\cite{BL92}) is
to specify the displacement $\xi^\mu$ in terms of the corresponding
infinitesimal geodesic. On this basis, according to (12) of
ref.~\cite{car93}, one finds that~\cite{car96} that the corresponding
Lagrangian value $\gL_{\mu\nu}$ of the perturbed metric will be given
in terms of its Eulerian analogue $\gE_{\mu\nu}$ and the associated
Riemannian differentiation operator $\nabE_{\!\mu}$ and curvature
$\RE_{\mu\nu\rho\sigma}$  by
\bel{Lagran}
\gL_{\mu\nu}-\gE_{\mu\nu}=2\nabE_{\!(\mu}\xi_{\nu)}+\big(\nabE_{\!\mu}
\xi^\rho\big)\nabE_{\!\nu}\xi_\rho-\xi^\rho\xi^\sigma
{\RE}{_{\mu\rho\nu\sigma}}+{\cal O}\{\delta^3\} \,.\feq
Subject to the understanding that the metric perturbation is at most
of the same infinitesimal order as the displacement,
$h_{\mu\nu}={\cal O}\{\delta\}$, it then follows
that the required Lagrangian variation of the metric will be
expressible in terms of the unperturbed Riemannian differentiation operator
$\nabla_{\!\mu}$ and curvature ${\cal R}_{\mu\nu\rho\sigma}$ by the 
ubiquitously applicable formula
\bel{Lagranpert}
\Delta_{\mu\nu}-h_{\mu\nu}=2\nabla_{\!(\mu}\xi_{\nu)}+\xi^\rho
\nabla_{\!\rho} h_{\mu\nu}+2h_{\rho(\mu}\nabla_{\!\nu)}\xi^\rho+\big(
\nabla_{\!\mu}\xi^\rho\big)\nabla_{\!\nu}\xi_\rho-\xi^\rho\xi^\sigma
{\cal R}_{\mu\rho\nu\sigma}+{\cal O}\{\delta^3\} \ .\feq
The linear order part given by the first term on the
right in this formula is the well known Lie derivative
term. The novelty here is inclusion of the quadratic order adjustment, whose
potential importance is to be emphasised: it is likely to be useful in more
general physical contexts, not just for treating the particular case of the
Dirac-Goto-Nambu strings and higher branes that are considered here. 

It is to be remarked that there is a lot of arbitrary gauge freedom in
the way one chooses the displacement generating field $\xi^\mu$ at
positions off the worldsheet. However, although derivatives of the
displacement field in directions orthogonal to the worldsheet may play
a role in the intermediate steps of the calculations, they must cancel
out (a condition which provides a useful check on the algebra) in  the
physical equations governing the motion of the worldsheet, for which
only the values of $\xi^\mu$ at positions actually on the worldsheet
are relevant. In the particular case to be considered here,
Dirac-Nambu-Goto branes, there is another kind of gauge freedom
involved, related to the choice of the tangential component of the
displacement fields. This is because moving along  the worldsheet has
no effect on its locus, which --- in the absence of internal currents
and fields --- is the only physical structure that matters. Hence, in
the Dirac-Goto-Nambu case there is no loss of physical generality if,
for the purpose of obtaining well behaved hyperbolic dynamical
equations, one chooses to eliminate this freedom by fixing the gauge on
the worldsheet through imposing the condition of orthogonality,
$\eta_{\mu\nu}\xi^\mu=0$.  In the present article, we shall
nevertheless impose no restriction on the gauge to be used, leaving
open the possibility of including an arbitrary tangential component in
the choice of $\xi^\mu$.

\section{Second order variation of the action.} 

We are now in a position to vary the action (\ref{action}) in the form
(\ref{expansion}). It will be convenient to use an abbreviation scheme
in which an acute accent indicates differentiation with respect to the
relevant variation parameter, so that in the particular case of the
action we are able to simply write $\delta \A = \AI $ and
$\delta^2\A=\AII$. We should note that, since the action is global, the
distinction between Lagrangian and Eulerian variations does not arise
at this stage.

It is evident from (\ref{Lagranpert}) that in this scheme the first order
Lagrangian variation of the metric  will be given by an expression of the
familiar  form
\bel{1stlag}
\dL g_{\mu\nu}= \hI_{\mu\nu}+ 2\nabla_{\!(\mu}\xiI_{\nu)} \feq
while the corresponding second order Lagrangian variation of the metric
will be given by
\bel{2ndlag}   \dL^2 g_{\mu\nu} =\hII_{\mu\nu}+ 
2\nabla_{\!(\mu}\xiII_{\nu)} + 2\xiI^\rho
\nabla_{\!\rho} \hI_{\mu\nu}+4\hI_{\rho(\mu}\nabla_{\!\nu)}
\xiI^\rho+2\big(\nabla_{\!\mu}\xiI^\rho\big)\nabla_{\!\nu}\xiI_\rho
-2\xiI^\rho\xiI^\sigma{\cal R}_{\mu\rho\nu\sigma}  \,.\feq
This last expression can be slightly simplified by ommission of the
first term on the right if the perturbing field $h_{\mu\nu}$ is
prescribed to scale directly with the variation parameter, so that one
simply has $h_{\mu\nu} = \hI_{\mu\nu}$ and $\hII_{\mu\nu}=0$. However
one can not analogously exclude the presence of higher order
contributions in the expansion $\xi^\mu=\xiI^\mu+ {_1\over
^2}\xiII^\mu+ ...$ for the a priori unknown, displacement field. At
this stage we do not make any simplification, but subsequent sections
we shall.

To allow for the effects of the perturbation on the surface measure, whose
first order Lagrangian variation will be given~\cite{batcar95} in terms of the
fundamental tensor $\eta^{\mu\nu}$  by an expression of the familiar form
\bel{delsigma}
\dL\big(d\ov{\mit\Sigma}\big) ={_1\over ^2}\eta^{\mu\nu}\big(
\dL g_{\mu\nu}\big)\, d\ov{\mit\Sigma} \,,\feq
it is convenient to express the first and second order variations of any 
action integral of the form (\ref{action}) as
\bel{d12act}
\AI=\int\big(\diamL\ov\La\,\big)\,d\ov{\mit\Sigma} \,, \hskip 1 cm
\AII=\int\big(\diamL^2\ov\La\,\big)\,d\ov{\mit\Sigma}\,.
\feq

The measure weighted `diamond-differential' operator
$\diamondsuit$~\cite{car94} is surface generalisation of the
`boldface-differential' operator already used by Friedman and
Schutz~\cite{frisc75}). When applied to the Lagrangian density it gives
\bel{diam1}
\diamL \ov\La=\dL \ov\La+{_1\over^2}\ov\La\,\eta^{\mu\nu}\,\dL 
g_{\mu\nu}\,,\feq
so that at second-order one obtains
\bel{diam2}
\diamL^2\ov\La=\dL^2\ov\La+\big(\dL\ov\La\big)
\eta^{\mu\nu}\dL g_{\mu\nu}+{_1\over^4}\ov\La
\big(\eta^{\mu\nu}\eta^{\rho\sigma}-2\eta^{\mu\rho}
\eta^{\nu\sigma}\big)(\dL g_{\mu\nu})\dL g_{\rho\sigma}+{_1\over^2}\ov\La\,
\dL^2 \eta^{\mu\nu}g_{\mu\nu}\,.\feq
Provided any internal fields that may be present
satisfy the corresponding dynamical equations, which implies that they give
no contribution to the first-order variation, and provided the only
relevant external field is that of the background metric $g_{\mu\nu}$,
then the variation in (\ref{d12act}) will be expressible
just in terms of the surface energy-momentum density
tensor $\ov T^{\mu\nu}$ in the standard form~\cite{batcar95}
\bel{stress} 
\delta{\cal I}=\,{_1\over^2} \int\ov T^{\mu\nu}\big(\dL
g_{\mu\nu} \big)\, d\ov{\mit\Sigma}\,,\hskip 1 cm \ov
T{^{\mu\nu}}=2{\partial \ov\La\over\partial g_{\mu\nu}}+
\ov\La\eta^{\mu\nu}\,.\feq
 
We shall now restrict our
attention to the simple case of a Dirac-Goto-Nambu membrane or string,
for which Lagrangian density in (\ref{action}) is just a constant,
\bel{DGNact} 
\ov\La=-m^{\rm p+1} \,,\feq
where $m$ is a fixed parameter which in natural units will have the
dimensions of mass. For topological defect system, such as a cosmic string or
domain wall, this Kibble mass
parameter will have the same order of magnitude as the Higgs mass associated
with the underlying spontaneous symmetry breaking. In this simple case, the
Lagrangian variation of the Lagrangian density will be given trivially by
$\dL\ov L=0 $ and the surface energy-momentum density tensor
(\ref{stress}) will simply be given in  terms of the fundamental tensor by the
proportionality relation
\bel{DGNstress}
\ov T{^{\mu\nu}}=-m^{\rm p+1}\eta^{\mu\nu}\,.\feq

Under these circumstances it can immediately be seen from the preceeding
formulae (\ref{1stlag}) and (\ref{diam1}) that the integrand of the first 
order variation of the action will be given by
\bel{DGNvar1}
-m^{\rm-(p+1)}\diamL\ov\La={_1\over^2}\eta^{\mu\nu}\dL g_{\mu\nu}
={_1\over ^2}\eta^{\mu\nu}\hI_{\mu\nu}+\ov\nabla_{\!\mu}\xiI^\mu\,.\feq
The content of the variation principle is that the surface integral of
this quantity must be made independent of the infinitesimal
displacement vector field $\xiI^\mu$. In other words the dynamical
equations are specified as the condition for all the displacement
dependent terms on the the right hand side of (\ref{DGNvar1}) to be
contained within a surface current divergence, that is, a current which
is tangential to the worldsheet.  The relevant term
$\ov\nabla_{\!\mu}\xiI^\mu$ would already have the form of a surface
current divergence as it stands, if the displacement $\xiI^\mu$
satisfied the worldsheet tangentiality condition $\xiI^\mu=\eta^\mu_{\
\nu}\xiI^\nu$. More generally, in the non trivial case for which it has
an orthogonal part, $\perp^{\!\mu}_{\,\nu} \xiI^\mu\not=0 $, the
formula $\ov\nabla_{\!\nu}\!\perp^{\!\nu}_{\,\mu}=-K_\mu$ can be used
to expand the term in question as
\bel{divterm}
\ov\nabla_{\!\mu}\xiI^\mu=\ov\nabla_{\!\mu}\big(\eta^\mu_{\ \nu}\xiI^\nu
\big)-K_\mu\xiI^\mu\,,\feq
where $K_\mu$ is the curvature vector, as defined by (\ref{trace}).
Since the first term in (\ref{divterm}) has the form of a surface current
divergence, and as such is irrelevant from the point of view of the variation
principle, we can replace (\ref{DGNvar1}) by the expression
\bel{DGNequiv}
-m^{\rm -(p+1)}\diamL{\,\atop\,}\ov\La\equiv{_1\over ^2}
\eta^{\mu\nu}\hI_{\mu\nu}-K_\mu\xiI^\mu\,,\feq
using the equivalence symbol $\equiv$ to indicate equality modulo divergences
that are removable by Green's theorem for a variation with compact support. At
this first order level the metric variation $\hI_{\mu\nu}$ cannot couple to
the displacement. It thus becomes evident that the content of the ensuing
Dirac Goto Nambu type field equation is simply the well known condition that
this vector $K_\mu$ should vanish.

We are now ready to move onto new ground by using (\ref{2ndlag}) and
(\ref{diam2}) to obtain the second-order analogue of (\ref{DGNvar1}) which
works out as
\begin{eqnarray}
-m^{\rm -(p+1)}\diamL^2\ov\La&=&\big({_1\over^4}\eta^{\mu\nu}\eta^{\rho\sigma}-
{_1\over^2}\eta^{\mu\rho}\eta^{\nu\sigma}\big)\hI_{\mu\nu}\hI_{\rho\sigma}
+\perp^{\!\rho}_{\,\sigma}\big(2\hI_\rho^{\ \nu}\ov\nabla_{\!\nu}\xiI^\sigma
+\eta^{\mu\nu}\xiI^\sigma\nabla_{\!\rho}\hI_{\mu\nu} \big) \nonumber
+\eta^{\rho\sigma}\ov\nabla_{\!\nu}\big(\hI_{\rho\sigma}\xiI^\nu\big)
 \\ &+&2\big(\ov\nabla_{\![\mu}\xiI^{\,\mu}\big)\ov\nabla_{\!\nu]}\xiI^{\,\nu}
+\perp^{\!\rho}_{\,\sigma}\big(\ov\nabla_{\!\mu}\xiI_{\rho}\big)
\ov\nabla^\mu\xiI^\sigma-\eta^{\mu\nu}{\cal R}_{\mu\rho\nu\sigma}\xiI^\rho
\xiI^\sigma+{_1\over^2}\eta^{\mu\nu}\hII_{\mu\nu}+
\ov\nabla_{\!\nu}\xiII^\nu\,.
\label{DGNvar2}
\end{eqnarray}
In the manner already demonstrated in the special case~\cite{car93} for which
the gravitational perturbation $h_{\mu\nu}$ is absent, the second variational
integrand (\ref{DGNvar2}) will be employable as the Lagrangian density of a
variational principle for the perturbed dynamical equations at first-order.
For this purpose, the only relevant terms are those having homogeneously
quadratic dependence on first-order variations. The remaining terms, that is
to say the last two terms which are homogeneously linear in second-order
variations, are of the same form as the linear contribution (\ref{DGNvar1}) in
the first-order contribution, which means that as consequence of the zero
order field equations their contribution to the action integral will just be a
constant. Since the present analysis is based on the treatment of the
gravitational perturbation as a given background field, the the first term in
(\ref{DGNvar2}) will also just be a constant as far as the variational
principle is concerned

\section{Derivation of the perturbed dynamical equations}

Therefore, we have argued that dropping the constant
contributions (from the first term in (\ref{DGNvar2}) and from the second-order variations, which are irrelevant for the application of the variation
principle, one can obtain a second-order action ${\cal I}_{\{2\}}$ say, (differing
from $\AII$ in (\ref{d12act}) only by the omitted constant) of the form
\bel{newac}
{\cal I}_{\{2\}}=\int\ov {\cal L}_{\{2\}}\,d\ov{\mit\Sigma}\, ,\feq
where the relevant second-order Lagrangian is given by
\begin{eqnarray}
-m^{\rm -(p+1)}\ov {\cal L}_{\{2\}}&=&
2\big(\ov\nabla_{\![\mu}\xiI^{\,\mu}\big)\ov\nabla_{\!\nu]}\xiI^{\,\nu}
+\perp^{\!\rho}_{\,\sigma}\big(\ov\nabla_{\!\mu}\xiI_{\rho}\big)
\ov\nabla^\mu\xiI^\sigma-\eta^{\mu\nu}{\cal R}_{\mu\rho\nu\sigma}\xiI^\rho
\xiI^\sigma \nonumber \\
&+&\perp^{\!\rho}_{\,\sigma}\big(2\hI_\rho^{\ \nu}\ov\nabla_{\!\nu}\xiI^\sigma
+\eta^{\mu\nu}\xiI^\sigma\nabla_{\!\rho}\hI_{\mu\nu}
\big) +\eta^{\rho\sigma}\ov\nabla_{\!\nu}\big(\hI_{\rho\sigma}\xiI^\nu\big)
\,.
\end{eqnarray}

To apply the variation principle to ${\cal I}_{\{2\}}$ we need to consider a process in which the linearized field variables themselves undergo 
independent infinitesimal `virtual' variations 
$\xiI^\mu\rightarrow\xiI^\mu+\delta\xiI^\mu$ and
$\hI_{\mu\nu}\rightarrow \hI_{\mu\nu}+\delta\hI_{\mu\nu}$.
Writing $\xiJ^\mu=\delta\xiI^\mu$ and $\hJ_{\mu\nu}=\delta
\hI_{\mu\nu}$, that is,  using a grave accent to indicate
differentiation with respect to the `virtual' variation parameter
(as distinct from differentiation with respect to the `real' or `physical'
variations for which the acute accent is used), then the ensuing action
differential will have the form
\bel{secondvar}
\delta {\cal I}_{\{2\}}=2
\int{\cal I}_{\{1,1\}}\,d\ov{\mit\Sigma}\,,\feq
where ${\cal I}_{\{1,1\}}$ is a symmetric bi-linear functional of the independent
`real' (acute accented) and `virtual' (grave accented) variations,
that is given by
\begin{eqnarray}
-m^{\rm -(p+1)}\ov {\cal L}_{\{1,1\}}&=&
2\big(\ov\nabla_{\![\mu}\xiI^{\,\mu}\big)\ov\nabla_{\!\nu]}\xiJ^{\,\nu}
+\perp^{\!\rho}_{\,\sigma}\big(\ov\nabla_{\!\mu}\xiI_{\rho}\big)
\ov\nabla^\mu\xiJ^\sigma-\eta^{\mu\nu}{\cal R}_{\mu\rho\nu\sigma}\xiI^\rho
\xiJ^\sigma \nonumber \\
&+&\perp^{\!\rho}_{\,\sigma}\big(\hI_\rho^{\ \nu}\ov\nabla_{\!\nu}\xiJ^\sigma
+\hJ_\rho^{\ \nu}\ov\nabla_{\!\nu}\xiI^\sigma\big)+{_1\over^2} 
\perp^{\!\rho}_{\,\sigma}\eta^{\mu\nu}\big(\xiI^\sigma\nabla_{\!\rho}
\hJ_{\mu\nu}+\xiJ^\sigma\nabla_{\!\rho}\hI_{\mu\nu}
\big) +{_1\over^2}\eta^{\rho\sigma}\ov\nabla_{\!\nu}
\big(\hI_{\rho\sigma}\xiJ^\nu+\hJ_{\rho\sigma}\xiI^\nu\big)
\,.
\end{eqnarray}
At the expense of sacrificing the manifest symmetry between `real' and `virtual' contributions, the next step is to
introduce linear functionals $Q=Q\{\xi\}$ and $F=F\{h\}$ depending
respectively on the displacement vector $\xi^\mu$ and the gravitational
perturbation $h_{\mu\nu}$, so as rewrite this in the more directly applicable
form                     
\bel{bilin}
 -m^{\rm -(p+1)}\ov {\cal L}_{\{1,1\}}   =
\xiJ^\mu\big(\FI_\mu-\QI_\mu)
+\xiI^\mu\FJ_\mu+\ov\nabla_{\!\mu}\UJI^\mu\,,\feq
where we have used the obvious notation 
\beq
\QI_\mu=Q_\mu\{\xiI\}\, ,\hskip 1 cm \FI_\mu=F_\mu\{\hI\}\, ,
\hskip 1 cm \FJ_\mu=F_\mu\{\hJ\}\,.\feq
The surface current vector $ \UJI^\mu$ is a symmetric bi-linear
functional of the two variations, $Q_{\sigma}$ is the dynamical functional for the surface perturbation, and $F_{\sigma}$ is the equivalent functional for the gravitational perturbation. These expressions are given by,
\begin{eqnarray} 
\UJI^\mu&=&\perp^{\!\rho}_{\,\sigma}
\xiJ^\sigma\ov\nabla{^\mu}\xiI_\rho+ \eta^\mu_{\ \rho}\big(
\xiJ^\rho\ov\nabla_{\!\nu}\xiI^\nu-\xiJ^\nu\ov\nabla_{\!\nu}\xiI^\rho\big)  
 \eta^\mu_{\ \nu}\perp^{\!\rho}_{\ \sigma}\big(\xiJ^\sigma
\hI_\rho^{\ \nu}+\xiI^\sigma\hJ_\rho^{\ \nu}\big)+{_1\over^2}\eta^\mu_{\ \nu}
\eta^\rho_{\ \sigma}\big(\xiJ^\nu\hI_\rho^{\ \sigma}+\xiI^\nu
\hJ_\rho^{\ \sigma}\big)\,, \\
Q_\sigma\{\xi\}&=&\perp_{\sigma\rho}\ov\nabla_{\!\mu}\ov\nabla^\mu\xi^\rho
-2K_{\rho\ \sigma}^{\ \mu}\ov\nabla_{\!\mu}\xi^\rho
+2K_{[\sigma}\ov\nabla_{\!\nu]}\xi^\nu
+\perp^\mu_{\ \sigma}\eta^\lambda_{\ \nu}{\cal R}_{\lambda\mu\ \rho}
^{\,\ \ \nu}\xi^\rho\,, \\
\label{forceterm}
F_\sigma\{h\}&=&\perp^{\!\rho}_{\,\sigma}\big({_1\over^2}\eta^{\mu\nu}
\nabla_{\!\rho}h_{\mu\nu}-\ov\nabla_{\!\nu}h_\rho^{\ \nu}\big)
+\big(K_{\nu\ \sigma}^{\ \rho} -\perp^{\!\rho}_{\,\sigma}K_\nu
-{_1\over^2}\eta^\rho_{\ \nu}K_\sigma\big)h_\rho^{\ \nu}\,.
\end{eqnarray}
One can see that the last functional is automatically worldsheet
orthogonal, that is, for an arbitrary perturbing field $h_{\mu\nu}$  \bel{orthof}
\eta^\sigma_{\,\rho}F_\sigma\{h\}=0\,,\feq
is an identity.

The requirement of the variation principle is that the surface integral
(\ref{secondvar}) of (\ref{bilin}) should be independent of the virtual
displacement $\xiJ^\mu$ for a variation with compact support. Since the term
involving the current $\UJI^\mu$ is a pure surface divergence, this requirement
is evidently equivalent just to the condition that for a given `real'
gravitational perturbation field $\hI_{\mu\nu}$ the corresponding `real
displacement vector $\xiI^\mu$ should satisfy a field equation of the form
\bel{perteq}
\QI_\sigma=\FI_\sigma\, .\feq
The validity of this field equation can be immediately confirmed by checking
that when the unperturbed, zero order dynamical equations $K_\rho=0$ are
satisfied, (\ref{perteq}) does indeed agree with the first-order perturbed
dynamical equations given by equation (21) of our preceeding
work~\cite{batcar95}. In particular, it can be verified that the backreaction
terms $F_{1\sigma}$ and $F_{2\sigma}$ that were defined there by equation (45)
add up to a total that matches the gravitational forcing term on the right of
(\ref{perteq}), that is, the linear functional defined here by (\ref{forceterm})
is expressible in terms of our previous notation as
$F_\sigma=F_{1\sigma}+F_{2\sigma}$. According to (41) of ref.~\cite{batcar95}, the precise physical interpretation of this is that the
quantity $\ov f_\sigma=-m^{\rm(p+1)}F_\sigma$ represents the gravitational
force density exerted on the brane by a perturbing field $h_{\mu\nu}$.

\section{The symplectic current}

Having confirmed that the homogeneous quadratic functional ${\cal I}_{\{2\}}$
given by (\ref{DGNvar2}) does indeed act as a Lagrangian for the linearly
perturbed field equations, we can exploit it for the purpose of derivation of
various kinds of conservation law. In case where the gravitational perturbation $h_{\mu\nu}$ it was shown that the current treatment yielded a Noether type identity. This expresses the fact that, although it is not manifest in the  (\ref{bilin}), this bilinear variation term is by construction
symmetric in the sense of being invariant under interchange of the two
independent, mathematically equivalent variations involved. This symmetry property is equivalently
expressible as the vanishing of the antisymmetric quantity obtained by taking
the difference between (\ref{bilin}) and its analogue obtained by swapping
the acute and grave accents. This identity which reduces, by the symmetry
property of partial differentiation, to the form  
\bel{simplecidy}
\nabla_{\!\mu}{\CIJ}^\mu=  \xiJ^\mu\QI_\mu-\xiI^\mu\QJ_\mu \,,\feq
in terms of a symplectic --- meaning antisymmetric bilinear --- surface current
$\CIJ^\mu\equiv\C^\mu\{\xiJ,\xiI\}$ given by $\CIJ=-\CJI= 
\UJI^\mu-\UIJ^\mu$ where $\UIJ^\mu$ is obtained from $\UJI^\mu$ by the
interchange of `real' and `virtual' variations. This gives
\bel{symplec}
\CIJ^\mu=\big(\eta^{\mu\nu}\!\perp_{\rho\sigma}\!
+\,2\eta^\mu_{[\rho}\eta^\nu_{\sigma]}\big)
\big(\xiJ^\rho\ov\nabla_{\!\nu}\xiI^\sigma
-\xiI^\rho\ov\nabla_{\!\nu}\xiJ^\sigma\big) \,,\feq
in which it is to be noticed that which the external gravitational 
perturbation field has cancelled out. Hence, the geometrically defined 
current functional $C^\mu$ thus retains exactly the same form, apart 
from the omission of the physical weighting factor $m^{\rm p+1}$ in its
definition, as the corresponding dimensionally weighted symplectic current
${\cal C}^\mu$ that was obtained in ref.\cite{car93}. The current (\ref{symplec}) can be rewritten in the form
\bel{sympler}
\CIJ^\mu=\perp_{\rho\sigma}\!
\big(\xiJ^\rho\ov\nabla{^\mu}\xiI^\sigma
-\xiI^\rho\ov\nabla{^\mu}\xiJ^\sigma\big)
+2\eta^\mu_{\,\rho}\ov\nabla_{\!\sigma}
\big(\xiJ^{\,[\rho}\xiI^{\sigma]}\big)
 \, .\feq
from which it can be seen that, for the case of worldsheet conserving
displacements, that is,  when $\xiI^\mu$ and $\xiJ^\mu$ are both tangential to the brane, the first term  will simply vanish while the second term will have the form of an exact surface divergence. This implies that that it will be trivially conserved as an identity, independently of any field equations that
may be satisfied~\cite{car92}.

Although it retains the same form, the symplectic current now no longer obeys the simple surface current conservation law that was obtained~\cite{car93} in the absence of gravitational perturbations. $\xiJ^\mu$ is postulated to satisfy a dynamical
equation of the same form as the first order perturbation equation
(\ref{perteq}) that must be satisfied by $\xiI^\mu$, allowing the dependence on the displacements to be eliminated from the right hand side of (\ref{simplecidy}), and hence one is left
with a relation of the form
\bel{symplecdiv}
\nabla_{\!\mu}{\CIJ}^\mu=  \xiJ^\mu\FI_\mu-\xiI^\mu\FJ_\mu \,,\feq
in which, generically, there remains a source term on the right in cases
where the forcing effect of gravitational wave perturbations is present.

By the orthogonality property (\ref{orthof}),
the trivial tangential parts of  $\xiI^\mu$ and $\xiJ^\mu$ make no contribution to
the right hand side of (\ref{symplecdiv}). To see how they affect the
left hand side, it is useful to rewrite the formula (\ref{sympler}) for
the current in the form
\bel{sympl}
\CIJ^\mu=\perp_{\rho\sigma}\!
\big(\xiJ^\rho\ov\nabla{^\mu}(\perp^{\!\sigma}_{\,\nu}\xiI^\nu)
-\xiI^\rho\ov\nabla{^\mu}(\perp^{\!\sigma}_{\,\nu}\xiJ^\nu)\big)
+2\eta^\mu_{\,\nu}K_\sigma\xiI^{\,[\nu}\xiJ^{\sigma]}+2\eta^\mu_{\,\nu}
\ov\nabla_{\!\sigma}\big(\eta^{[\nu}_{\ \rho}\eta^{\sigma]}_{\ \tau}
\xiJ^{\,[\rho}\xiI^{\tau]}\big)
 \, ,\feq
in which the last term has the form of an exact surface divergence, so that
independently of the field equations it will be trivially conserved as an
identity. Provided the unperturbed field equations, $K_\mu=0$
are satisfied, the second term simply drops out, so the `reduced'
symplectic current vector $\CrIJ{^\mu}$ consisting just of the 
first term, namely
\bel{redsympl}
 \CrIJ{^\mu}=\perp_{\rho\sigma}\!
\big(\xiJ^\rho\ov\nabla{^\mu}(\perp^{\!\sigma}_{\,\nu}\xiI^\nu)
-\xiI^\rho\ov\nabla{^\mu}(\perp^{\!\sigma}_{\,\nu}\xiJ^\nu)\big)
\,,\feq
will satisfy an equation of the same form as the original
symplectic current, that is,
\bel{redsymplecdiv}
\nabla_{\!\mu} \CrIJ{^\mu}=  \xiJ^\mu\FI_\mu-\xiI^\mu\FJ_\mu \,,\feq
Unlike the original current ${\CIJ}{^\mu}$, this `reduced' current
$\CrIJ{^\mu}$ has the advantage of being gauge independent in the sense of
being entirely independent of the tangentially projected parts of the vectors
$\xiI^\mu$ and $\xiJ^\mu$ on which it depends.

In the kind of practical application for which such a Noetherian identity is
typically used, namely for exploiting underlying symmetries of various kinds,
it is usually sufficient to consider only virtual displacements satisfying the
unforced dynamical equations, which means taking $\hJ_{\mu\nu}=0$, and thereby
setting $\FJ_\mu=0$ in the right hand side of (\ref{symplecdiv}). A virtual
displacement field satisfying this condition will be given automatically by
setting $\xiJ^\mu= k^\mu$ where $k^\mu$ is any solution of the background
Killing equations $\nabla_{(\mu}k_{\nu)}=0$, since the action of such a
translation on any given unperturbed solution evidently translates it onto a
nearby solution that is geometrically identical and therefore also a solution.
What this means -- for any real physical solution $\xiI^\mu$ of the
displacement perturbation equation (\ref{perteq}) -- is that associated with
the geometric symmetry generated by $k^\mu$ there will be a corresponding
generalised momentum current
\beq
\CrI{^\mu}\{k\}=\perp_{\rho\sigma}\!
\big(k^\rho\ov\nabla{^\mu}(\perp^{\!\sigma}_{\,\nu}\xiI^\nu)
-\xiI^\rho\ov\nabla{^\mu}(\perp^{\!\sigma}_{\,\nu}k^\nu)\big)
\,,\feq
that can be seen from (\ref{symplecdiv}) to obey a divergence equation of 
the form
\bel{diveq}
\nabla_{\!\mu}\CrI\{k\}^\mu=  k^\mu\FI_\mu
 \,.\feq

In view of the orthogonality property (\ref{orthof}) the source term on the
right will evidently vanish in the case for which $k^\mu$ is tangent to the
worldsheet, $\perp^{\!\mu}_{\,\nu}\!k^\nu=0$, i.e. when the {\it unperturbed
solution is invariant} under the action of the relevant symmetry,  but in
such a case the ensuing strict conservation law is vacuous, since the current
$\CrI\{k\}^\mu$ will itself be zero. In the case for which the 
unperturbed solution is invariant under the action of $k^\mu$
one can however obtain a non trivial application of (\ref{redsymplecdiv}) by
using the fact that in these circumstances if the first order dynamical
equation (\ref{perteq}) is satisfied by a ``real'' displacement  field
$\xiI^\mu$ for a given `real'' gravitational perturbing field
$\hI_{\rho\sigma}$, then the first order dynamical equation will also be
satisfied by the Lie derivatives of these fields with respect to $k^\mu$, so
these Lie derivatives will be utilisable used as the ``virtual'' fields in
(\ref{redsymplecdiv}), i.e. we shall be able to take
$\xiJ^\mu=$ $k^\nu\nabl_{\!\nu}\xiJ^\mu-\xiI^\nu\nabla_{\!\nu}k^\nu$ and
$\hJ_{\rho\sigma}=$ $k^\nu\nabla_{\!\nu} \hI_{\rho\sigma}
+2\hI_{\nu(\sigma}\nabla_{\!\rho)}k^\nu$.

\bigskip
\noindent {\bf ACKNOWLEDGEMENTS}
\medskip

\noindent We wish to thank Luc Blanchet and Paul Shellard for helpful conversations. RAB is funded by Trinity College.


\end{document}